\documentclass[aps,prd,twocolumn,a4paper]{revtex4}
\usepackage{ulem}
\usepackage{color}
\usepackage{graphicx}
\usepackage{epsfig}

\def\nn {\nonumber}
\newcommand{\be}{\begin{equation}}
\newcommand{\ee}{\end{equation}}
\newcommand{\bea}{\begin{eqnarray}}
\newcommand{\eea}{\end{eqnarray}}

\newcommand{\ep}{\epsilon}

\newcommand{\om}{\omega}

\newcommand{\ov}{\overline}

\newcommand{\qs}{q \!\!\! /}
\newcommand{\ks}{k \!\!\! /}

\newcommand{\ls}{l \!\!\! /}

\newcommand{\tom}{\widetilde{\om}}   

\newcommand{\vk}{\vec k}
\newcommand{\vp}{\vec p}
\newcommand{\vq}{\vec q}

\newcommand{\mn}{\mu\nu}
\newcommand{\del}{\partial}

\newcommand{\unit}{1\!\!1}

\begin{document}
\title{The nucleon thermal width due to pion-baryon loops and its contribution in Shear viscosity}
\author{Sabyasachi Ghosh}
\affiliation{Instituto de Fisica Teorica, Universidade Estadual Paulista, 
Rua Dr. Bento Teobaldo Ferraz, 271, 01140-070 Sao Paulo, 
SP, Brazil.}

\begin{abstract}
In the real-time thermal field theory, the standard expression
of shear viscosity for the nucleonic constituents is derived from the two point
function of nucleonic viscous stress tensors at finite temperature and density.
The finite thermal width or Landau damping is traditionally
included in the nucleon propagators. 
This thermal width is calculated from the in-medium self-energy
of nucleon for different possible pion-baryon loops. The dynamical
part of nucleon-pion-baryon interactions are taken care by the effective
Lagrangian densities of standard hadronic model.
The shear viscosity to entropy density ratio of nucleonic component
decreases with the temperature and increases with the nucleon chemical
potential. However, adding the contribution of pionic component,
total viscosity to entropy density ratio also reduces with the nucleon 
chemical potential when the mixing effect between pion and 
nucleon components in the mixed gas is considered.  
Within the hadronic domain, viscosity to entropy density ratio
of the nuclear matter is gradually reducing as temperature
and nucleon chemical potential are growing up 
and therefore the nuclear matter is approaching toward the (nearly) perfect fluid nature.
\end{abstract}


\maketitle
\section{Introduction}
The recent hydrodynamical~\cite{Romatschke,Heinz}
as well as some transport studies~\cite{Xu,Greco} 
have indicated about an (nearly) ideal fluid nature of nuclear matter,
which may be produced in the experiments of heavy ion collisions (HIC)
like Relativistic Heavy Ion Collider (RHIC) at BNL.
The hydrodynamical calculations became very successful
in explaining the elliptical flow parameter, $v_2$ from RHIC data~\cite{PHENIX,STAR,PHOBOS} 
only when they assumed a very small ratio of shear viscosity
to entropy density $(\eta/s)$ for the expanding nuclear matter.
When some recent studies~\cite{Csernai,Purnendu,Gyulassy,Hufner} 
(see also Ref.~\cite{Chen_Tc}) show that $\eta/s$ may reach a minimum in the 
vicinity of a phase transition, then some special attentions are drawn 
to the smallness of this minimum value with respect to its
lower bound ($\eta/s=\frac{1}{4\pi}$),
commonly known as the KSS bound~\cite{KSS}.
In this context, the temperature ($T$) dependence of $\eta/s$ 
is taken into account in some recent hydrodynamical 
calculations~\cite{Niemi,Heinz_T,Bhat,Krein} instead of
its constant value during the entire evolution.
Niemi {\it et al.}~\cite{Niemi} have interestingly observed that 
the $v_2(p_T)$ of RHIC data 
is highly sensitive to the temperature dependent $\eta/s$ 
in hadronic matter and almost independent of the viscosity in QGP phase.
This work gives an additional boost to the microscopic calculations 
of $\eta/s$ of the hadronic matter in the recent 
years~\cite{Prakash_2012,Greiner,Toneev,Buballa,Dobado,Nakano,Itakura,
Muronga,Nicola,Weise,SSS,SPal,Gorenstein,Denicol,Bass,Fang,Sadooghi},
though historically these investigations are slightly 
old~\cite{Gavin,Prakash,Toneev_7,Toneev_8,Toneev_9,Toneev_10}. 

Except a few~\cite{Itakura,Gorenstein,Denicol,Bass},
most of the microscopic calculations are done
in zero baryon or nucleon chemical potential ($\mu_N=0$).
Along with the $T$ dependence of $\eta$ or $\eta/s$, their dependence on 
the baryon chemical potential should also be understood in view of
the future experiments such as FAIR. In the work of Itakura {\it et al.}~\cite{Itakura}
and Denicol {\it et al.}~\cite{Denicol}, we notice that 
the $\eta/s$ is reduced at finite baryon chemical potential,
whereas Gorenstein {\it et al.}~\cite{Gorenstein} observed an 
increasing nature of $\eta/s$
with $\mu_N$. Itakura {\it et al.} have obtained $\eta$ by solving the 
relativistic quantum Boltzmann equation, where phenomenological
amplitudes of hadrons are used in the collision terms.
Denicol {\it et al.} have calculated the $\eta$ at finite $T$ and $\mu_N$ 
by applying Chapman-Enskog theory in Hadron Resonance Gas (HRG) model,
whereas Gorenstein {\it et al.} have taken a simplified ansatz of $\eta$
to estimate $\eta/s$ in the van der Waals excluded volume HRG model.
Similar to the ansatz of $\eta(T)$ taken by Gorenstein {\it et al.}, 
the $\eta$ itself increases with increasing temperature in Ref.~\cite{Itakura},
but their $\eta/s$ are exhibiting completely opposite nature of $T$ dependence.
Therefore, the behavior of the $\eta/s$ may largely be influenced by
the $T$ dependence of entropy density $s$.

Motivating by these delicate issues of shear 
viscosity at finite $\mu_N$, the present manuscript is
concentrated on the matter with nucleon degrees of freedom at finite
$T$ and $\mu_N$. The nucleons in the medium can slightly
become off-equilibrium because of their thermal width or 
Landau damping, which can be originated from the nucleon thermal
fluctuations into different baryons and pion. The inverse
of nucleon thermal width measures the relaxation time of nucleon in the 
matter from which one can estimate its corresponding 
shear viscosity contribution.

In the next section, the one-loop expression
of $\eta$ for nucleon degrees of freedom is derived from the Kubo relation,
where a finite thermal width is traditionally included
in the nucleon propagators. This standard expression of $\eta$
can also be deduced from relaxation time approximation of kinematic
theory approach. In the real-time thermal field theory,
the nucleon thermal width
from the different pion-baryon loops are calculated in Sec.~3,
where their interactions are determined from the effective hadronic model.
In Sec.~4, the numerical results are discussed
followed by summary and conclusions in Sec.~5.

\section{Kubo relation for shear viscosity of nuclear matter}
From the simple derivation of Kubo formula~\cite{Zubarev,Kubo},
let us start with the expression of shear viscosity
for nucleonic constituents in momentum space~\cite{G_Kubo,Nicola},
\be
\eta_N=\frac{1}{20}\lim_{q_0,\vq \rightarrow 0}\frac{A_\eta(q_0,\vq)}{q_0}~,
\label{eta_Nicola}
\ee
where 
\be
A_\eta(q_0,\vq)=\int d^4x e^{iq\cdot x}\langle[\pi_{\mn}(x),\pi^{\mn}(0)]\rangle_\beta
\ee
is the spectral representation of two point function for nucleonic viscous-stress tensor, 
$\pi^{\mn}$ and
\be
\langle \hat{O}\rangle_\beta={\rm Tr}\frac{e^{-\beta H}\hat{O}}{Z}
~~~~{\rm with}~~ Z={\rm Tr}e^{-\beta H}
\ee
is denoting the thermodynamical ensemble average.
The energy momentum tensor of free nucleon is
\bea
T_{\rho\sigma}&=&-g_{\rho\sigma}{\cal L}
+\frac{\del {\cal L}}{\del(\del^\rho\psi)}\del_\sigma\psi
\nn\\
&=&-g_{\rho\sigma}{\cal L}+i{\ov \psi}\gamma_\rho\del_\sigma\psi~,
\eea
and hence the viscous stress tensor will be
\bea
\pi_{\mn}&=&t^{\rho\sigma}_{\mn}T_{\rho\sigma}
\nn\\
&=&t^{\rho\sigma}_{\mn}i{\ov \psi}\gamma_\rho\del_\sigma\psi
~~(~{\rm since} ~t^{\rho\sigma}_{\mn}g_{\rho\sigma}=0~)~,
\label{vis_stress}
\eea
where 
\be
t^{\rho\sigma}_{\mn}=\Delta^\rho_\mu\Delta^\sigma_\nu
-\frac{1}{3}\Delta_{\mn}\Delta^{\rho\sigma},~~
\Delta^{\mn}=g^{\mn}-u^\mu u^\nu~.
\ee
In real-time formalism of thermal field theory, the ensemble average of any two point function
always becomes a $2\times 2$ matrix structure. Hence, for viscous-stress tensor, 
the matrix structure of two point function becomes
\be
\Pi_{ab}(q)=i\int d^4x e^{iqx}\langle T_c\pi_{\mn}(x)\pi^{\mn}(0)\rangle^{ab}_\beta~,
\label{pi_ab}
\ee
where the superscripts $a, b (a,b=1,2)$ denote the thermal indices
of the matrix and $T_c$ denotes time ordering with respect to a 
symmetrical contour~\cite{Semenoff,SS_SM}
in the complex time plane.

\begin{figure}
\begin{center}
\includegraphics[scale=0.45]{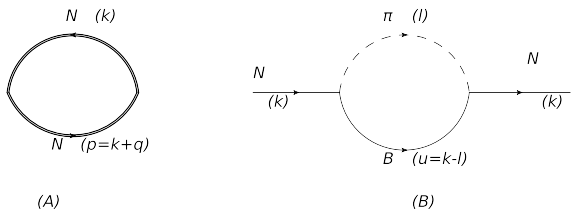}
\caption{Diagrammatic representation of $NN$ loop is shown in (A), where
double lines stand for effective $N$ propagators, which contain their
thermal widths $\Gamma$. The diagrammatic representation of 
nucleon self-energy for $\pi B$ loop is shown
in (B) from where $\Gamma$ can be determined.}
\label{Shear_N}
\end{center}
\end{figure}

The matrix can be diagonalized in terms of 
a single analytic function, which can also be related
with the retarded two point function of viscous-stress tensor.
The retarded function $\Pi^R(q)$, diagonal element ${\ov \Pi}(q)$
and the spectral function $A_\eta(q)$
are simply related to any one of the components of $\Pi_{ab}(q)$.
Their relations with 11 component is given below
\bea
A_\eta(q)&=&2{\rm Im}\Pi^R(q)=2\epsilon(q_0){\rm Im}{\ov \Pi}(q)
\nn\\
&=&2{\rm tanh}(\frac{\beta q_0}{2}){\rm Im}\Pi_{11}(q)~.
\label{R_bar_11}
\eea
Hence, Eq.~(\ref{eta_Nicola}) can broadly be redefined as 
\bea
\eta_N&=&\frac{1}{10}\lim_{q_0,\vq \rightarrow 0}
\frac{{\rm Im}\Pi^R(q_0,\vq)}{q_0}=\frac{1}{10}\lim_{q_0,\vq \rightarrow 0}
\frac{{\ep(q_0)\rm Im}{\ov \Pi}(q_0,\vq)}{q_0}
\nn\\
&=&\frac{1}{10}\lim_{q_0,\vq \rightarrow 0}
\frac{{\rm tanh}(\beta q_0/2){\rm Im}\Pi^{11}(q_0,\vq)}{q_0}~.
\label{eta_Im_R}
\eea
Using (\ref{vis_stress}) in the 11 component of (\ref{pi_ab}) and then
applying the Wick's contraction technique, we have
\bea
\Pi_{11}(q)&=&t^{\rho\sigma}_{\alpha\beta}t^{\alpha\beta}_{\mn}i\int d^4x e^{iqx}
\langle T{\ov \psi}\underbrace{(x)\gamma_\rho\del_\sigma\psi\overbrace{(x)
{\ov \psi}}(0)\gamma^\mu\del^\nu\psi}(0)\rangle_\beta
\nn\\
&=&i\int \frac{d^4k}{(2\pi)^4}N(q,k) 
D_{11}(k)D_{11}(p=q+k)~,
\label{P_11}
\eea
where 
\be
N(q,k)=-I_Nt^{\rho\sigma}_{\mn}{\rm Tr}[\gamma^\mu(q+k)^\nu(\qs+\ks+m_N)
\gamma_\rho k_\sigma(\ks+m_N)]~.
\ee
This self-energy function, $\Pi_{11}(q)$ for $NN$ loop can
diagrammatically be represented by Fig.~\ref{Shear_N}(A). 
In the co-moving frame, {\it i.e.}, for $u=(1,{\vec 0})$, the $N(q,k)$ becomes
\bea
N(q,k)&=&-I_N\left[\frac{32}{3}\{k_0(q_0+k_0)\}\{\vk\cdot(\vq+\vk)\}
\right.\nn\\
&&\left.-4\left[\{\vk\cdot(\vq+\vk)\}^2+\frac{\vk^2(\vq+\vk)^2}{3}\right]\right]~.
\label{N_qk}
\eea
In the above equations, $I_N=2$ is the isospin degeneracy of nucleon.

In Eq.~(\ref{P_11}), $D^{11}$ is scalar part of 11 component
of the nucleon propagator at finite temperature and density.
Its form is
\bea
&&D^{11}(k)=\frac{-1}{k^2-m_N^2+i\eta}-2\pi i
F_k(k_0)\delta(k^2-m_N^2)
\nn\\
&&~{\rm with}~~ F_k(k_0)=n^+_k\theta(k_0)+
n^-_k\theta(-k_0)\nonumber\\
\nn\\
&&~~=-\frac{1}{2\om^N_k}(\frac{1-n^+_k}{k_0-\om^N_k+i\eta}+
\frac{n^+_k}{k_0-\om^N_k-i\eta}
\nn\\
&&~~-\frac{1-n^-_k}{k_0+\om^N_k-i\eta}
-\frac{n^-_k}{k_0+\om^N_k+i\eta}),
\label{de11}
\eea
where $n^{\pm}_k(\om^N_k)=1/\{e^{\beta(\om^N_k \mp \mu_N)}+1\}$
is Fermi-Dirac distribution function for energy $\om^N_k=\sqrt{\vk^2+m_N^2}$. 
Here the $\pm$ signs in the superscript of $n_k$ stand for nucleon and 
anti-nucleon respectively.  
Among the four terms in Eq.~(\ref{de11}), the first and the second terms  
are associated with the nucleon propagation
above the Fermi sea and the propagation of its hole in the Fermi sea 
respectively, while the third and
fourth terms represent the corresponding situations for
anti-nucleon. 
The full relativistic nucleon propagator, thus, treats the
particle and anti-particle on an equal footing and all possible singularities 
(nucleon, hole of the nucleon, anti-nucleon and hole of the anti-nucleon) 
are automatically included. 

After doing the $k_0$ integration of Eq.~(\ref{P_11}) and then using it
in Eq.~(\ref{eta_Im_R}), we have
\bea
\eta_N&=&\frac{1}{10}\lim_{q_0,\vq \rightarrow 0}\int\frac{d^3k}{(2\pi)^3}
\frac{(-\pi N)}{4\om^N_k\om^N_p}
\nn\\
&&\left[\frac{\{-n^-_k(\om^N_k)+n^-_p(-q_0+\om^N_k)\}}{q_0}\delta(q_0 -\om^N_k+\om^N_p)
\right.\nn\\
&&\left.+\frac{\{n^+_k(\om^N_k)-n^+_p(q_0+\om^N_k)\}}{q_0}\delta(q_0 +\om^N_k-\om^N_p)
+..\right]~,
\nn\\
\label{Pi_LU}
\eea
where $N=N(k_0=\pm\om^N_k,\vk,q)$
and $\om^N_p=\sqrt{(\vq+\vk)^2+m_N^2}$.

The two $\delta$-functions will be responsible for
generating the Landau cuts ($-\vq<q_0<\vq$), where
the Im$\Pi^R(q)$ will be non-zero. However, there will be
two more $\delta$-functions (not written explicitly), which
are not important for the limiting point $q_0, \vq\rightarrow 0$ since they
will generate unitary cuts ($-\infty<q_0<-\sqrt{\vq^2+4m_N^2}$
and $\sqrt{\vq^2+4m_N^2}<q_0<\infty$).

Using the identity 
\be
-\pi\delta(x)={\rm Im}\left[\lim_{\Gamma_N \rightarrow 0}
\frac{1}{x+i\Gamma_N}\right]
\ee
in
Eq.~(\ref{Pi_LU}), we have
\bea
\eta_N
&=&\frac{1}{10}\lim_{q_0,\vq \rightarrow 0}{\rm Im}
\left[\int\frac{d^3k}{(2\pi)^3}\frac{N}{4\om^N_k\om^N_p}\lim_{\Gamma_N \rightarrow 0}
\right.\nn\\
&&\left.\left\{\frac{\{-n^-_k(\om^N_k)+n^-_p(-q_0+\om^N_k)\}/q_0}{(q_0-\om^N_k+\om^N_p)+i\Gamma_N}
\right.\right.\nn\\
&&\left.\left.+\frac{\{n^+_k(\om^N_k)-n^+_p(q_0+\om^N_k)\}/q_0}{(q_0+\om^N_k-\om^N_p)+i\Gamma_N}\right\}\right]~.
\label{eta_Gama}
\eea
We will continue our further calculation
for finite value of $\Gamma_N$ to get a non-divergent
contribution of $\eta_N$. Including thermal width $\Gamma_N$
for constituent particles (here nucleons) of the medium is a 
very well established technique~\cite{Hosoya,Nicola,Weise} 
in Kubo approach to remove
the divergence of $\eta_N$ as well as to incorporate the 
interaction scenario, which is
very essential for a dissipative system. The interaction scenario
is coming into the picture
by transforming the delta functions to the spectral
functions with finite thermal width. 
The thermal width (or collision rate) $\Gamma_N$
of the constituent particles reciprocally measures the shear 
viscosity coefficient, which is approximately equivalent to the  
quasi particle description.

In the limiting case of $q_0,\vq\rightarrow 0$, we get
$\om^N_p\rightarrow\om^N_k$ and
therefore Eq.~(\ref{eta_Gama}) is transformed to
\be
\eta_N=\frac{1}{10}\int\frac{d^3k}{(2\pi)^3}\frac{(-N_0)}{4{\om^N_k}^2\Gamma_N}
\left[I_2+I_3\right]~,
\label{eta_I}
\ee
where 
\be
N_0=\lim_{q_0, \vq\rightarrow 0}N(k_0=\pm\om^N_k,\vk,q)
\label{N_0}
\ee
and
\be
I_{2,3}=\lim_{q_0 \rightarrow 0}
\frac{\{\mp n^\mp_k(\om^N_k)\pm n^\mp_p(\mp q_0+\om^N_k)\}}{q_0}~.
\label{I_23}
\ee
In the above Eq.~(\ref{I_23}), one can notice that the limiting value
of $I_{2,3}$ is of the $0/0$ form.
Therefore, we can apply the L'Hospital's rule, {\it i.e.}, 
\bea
I_{2,3}&=&\lim_{q_0\rightarrow 0}\frac{\frac{d}{dq_0}
\{\mp n^\mp_k(\om^N_k)\pm n^\mp_p(\mp q_0+\om^N_k)\}}{\frac{d}{dq_0}\{q_0\}}
\nn\\
&=&\beta [n^{\mp}_k(1-n^{\mp}_k)]~,
\eea
since
\bea
\frac{d}{dq_0}\{\pm n^{\mp}_p(\om_q=\mp q_0+\om^N_k)\}&=&\pm\frac{-\beta \frac{d\om_q}{dq_0} 
e^{\beta(\om_q\pm\mu_N)}}{\{e^{\beta(\om_q\pm\mu_N)}+1\}^2}
\nn\\
\lim_{q_0\rightarrow 0}\frac{d}{dq_0}\{\pm n^{\mp}_p(\om_q=\mp q_0+\om^N_k)\}
&=&\pm\frac{-(\mp)\beta 
 e^{\beta(\om^N_k\pm\mu_N)}}{\{e^{\beta(\om^N_k\pm\mu_N)}+1\}^2}
\nn\\
&=&\beta [n^{\mp}_k(1-n^{\mp}_k)]~. 
\eea
Again, in the limiting value of $q_0,\vq\rightarrow 0$,
Eq.~(\ref{N_qk}) can be simplified to
\be
N^0=-I_N\frac{16\vk^4}{3}~.
\ee
Hence, using the above results, the Eq.~(\ref{eta_I}) becomes
\bea
\eta_N&=&\frac{8\beta I_N}{15}\int\frac{d^3k}{(2\pi)^3}\frac{\vk^4}{4{\om^N_k}^2\Gamma_N}
\left[n^-_k(1-n^-_k)+n^+_k(1-n^+_k)\right]
\nn\\
&=&\frac{\beta I_N}{15\pi^2}\int \frac{\vk^6d\vk}{{\om^N_k}^2\Gamma_N}[n^-_k(1-n^-_k)
+n^+_k(1-n^+_k)]~.
\label{eta_last}
\eea
This is the one-loop expression of shear viscosity for the matter
with nucleon degrees of freedom in the
Kubo approach. 
Though there are possibility of infinite number of ladder-type diagrams, which
are supposed to be of same order of magnitude (${\cal O}(1/\Gamma_N)$) like the one-loop,
they will be highly suppressed~\cite{G_Kubo}.
As we increase the number of loops, the number of extra thermal 
distribution functions will also appear in the shear viscosity
expression and hence their numerical suppression will successively grow.
On this basis, the one-loop results may be considered as a leading order results.
One can derive exactly same expression from relaxation
time approximation in kinetic theory approach. 

\section{Calculation of nucleon thermal width}
\begin{table}  
\begin{center}
\label{tab1}
\begin{tabular}{|c|c|c|c|c|c|}
\hline
& & & & & \\
Baryons & $J_B^P$ & $I_B$ & $\Gamma_{\rm tot}$ & $\Gamma_{B\rightarrow N\pi}$ (B.R.) & $f/m_\pi$ \\
& & & & &  \\
\hline
& & & & & \\
$\Delta^(1232)$ & ${\frac{3}{2}}^+$ & 3/2 & 0.117 & 0.117 (100\%) & 15.7 \\
& & & & & \\
$N^*(1440)$ & ${\frac{1}{2}}^+$ & 1/2 & 0.300 & 0.195 (65\%) & 2.5 \\
& & & & & \\
$N^*(1520)$ & ${\frac{3}{2}}^-$ & 1/2 & 0.115 & 0.069 (60\%) & 11.6 \\
& & & & & \\
$N^*(1535)$ & ${\frac{1}{2}}^-$ & 1/2 & 0.150 & 0.068 (45\%) & 1.14 \\
& & & & & \\
$\Delta^*(1600)$ & ${\frac{3}{2}}^+$ & 3/2 & 0.320 & 0.054 (17\%) & 3.4 \\
& & & & & \\
$\Delta^*(1620)$ & ${\frac{1}{2}}^-$ & 3/2 & 0.140 & 0.035 (25\%) & 1.22 \\
& & & & & \\
$N^*(1650)$ & ${\frac{1}{2}}^-$ & 1/2 & 0.150 & 0.105 (70\%) & 1.14 \\
& & & & & \\
$\Delta^*(1700)$ & ${\frac{3}{2}}^-$ & 3/2 & 0.300 & 0.045 (15\%) & 9.5 \\
& & & & & \\
$N^*(1700)$ & ${\frac{3}{2}}^-$ & 1/2 & 0.100 & 0.012 (12\%) & 2.8 \\
& & & & & \\
$N^*(1710)$ & ${\frac{1}{2}}^+$ & 1/2 & 0.100 & 0.012 (12\%) & 0.35 \\
& & & & & \\
$N^*(1720)$ & ${\frac{3}{2}}^+$ & 1/2 & 0.250 & 0.028 (11\%) & 1.18 \\
& & & & & \\
\hline
\end{tabular}
\caption{From the left to right columns, the table contain
the baryons, their spin-parity quantum numbers $J_B^P$,
isospin $I_B$, total decay width $\Gamma_{\rm tot}$,
decay width in $N\pi$ channels $\Gamma_{B\rightarrow N\pi}$ or
$\Gamma_B(m_B)$ in Eq.~(\ref{Gam_BNpi}) (brackets displaying 
its Branching Ratio) and at the last coupling constants $f/m_\pi$.}
\end{center}
\end{table}
Now, our next aim is to calculate the thermal width of nucleon $\Gamma_N$,
which can be estimated from
the retarded component of nucleon self-energy ($\Sigma^R$) at finite temperature
and density. Their relation is given by
\be
\Gamma_N(\vk,T,\mu_N)=-{\rm Im}\Sigma^R(k_0=\om^N_k,\vk,T,\mu_N)~.
\label{Gam_F}
\ee
During the propagation in the hot and dense nuclear matter,
nucleon may pass through
different $\pi B$ loops, where $B$ stand for different
higher mass baryons including nucleon itself. In this work, all possible
4-star baryon resonances with spin $1/2$ and $3/2$
are considered. These are $N(980)$, $\Delta(1232)$, $N^*(1440)$, $N^*(1520)$,
$N^*(1535)$, $\Delta^*(1600)$, $\Delta^*(1620)$, $N^*(1650)$, 
$\Delta^*(1700)$, $N^*(1700)$, $N^*(1710)$ and $N^*(1720)$,
where masses (in MeV) of the baryons are given inside the brackets.
The nucleon self-energy for $\pi B$ loop is shown in 
diagram~\ref{Shear_N}(B) and its 11 component can be expressed as
\bea
\Sigma^{11}(k,T,\mu_N)&=&-i\int \frac{d^4l}{(2\pi)^4}L(k,l)D_{11}(l,m_\pi,T)
\nn\\
&&D_{11}(u=k-l,m_B,T,\mu_N)~,
\label{Pi_11k}
\eea
where $D_{11}(l,m_\pi,T)$, $D_{11}(u=k-l,m_B,T,\mu_N)$ are scalar part of pion and
baryon propagators at finite temperature and density. The $L(k,l)$ contains
vertices and numerator parts of the propagators.
The chemical potential of all baryons are assumed to be the same
as nucleon chemical potential $\mu_N$.
Similar to Eq.~(\ref{R_bar_11}), this 11 component is also related with
retarded component as
\be
{\rm Im}\Sigma^R(k)={\rm coth}\left\{\frac{\beta (k_0-\mu_N)}{2}
\right\}{\rm Im}\Sigma_{11}(k)~.
\label{R_bar_N}
\ee
Performing the $l_0$ integration in (\ref{Pi_11k}) and then using
the relation (\ref{R_bar_N}), we get the imaginary part of retarded self-energy,
\bea
{\rm Im}{\Sigma}^{R}(k)&=&\pi\int\frac{d^3l}{(2\pi)^3}\frac{1}{4\om^\pi_l\om^B_u}
[L(l_0=\om^\pi_l,{\vec l},k)
\nn\\
&&[\{1+n_l(\om^\pi_l)-n^+_u(k_0-\om^\pi_l)\}\delta(k_0 -\om^\pi_l-\om^B_u)
\nn\\
&&+\{-n_l(\om^\pi_l)-n^-_u(-k_0+\om^\pi_l)\}\delta(k_0-\om^\pi_l+\om^B_u)]
\nn\\
&&+L(l_0=-\om^\pi_l,{\vec l},k)[\{n_l(\om^\pi_l)
\nn\\
&&+n^+_u(k_0+\om^\pi_l)\}\delta(k_0 +\om^\pi_l-\om^B_u)+\{-1
\nn\\
&&-n_l(\om^\pi_l)+n^-_u(-k_0-\om^\pi_l)\}\delta(k_0 +\om^\pi_l+\om^B_u)]]~,
\nn\\
\label{self_LU}
\eea
where $\om^B_u=\sqrt{({\vk}-{\vec l})^2+m_B^2}$,
$n^\pm_u$ and $n_l$ are respectively Fermi-Dirac and 
Bose-Einstein distribution functions. The regions of 
different branch cuts in $k_0$-axis are
$(-\infty$ to $-\sqrt{\vk^2+(m_\pi+m_B)^2}~)$ for unitary cut in negative $k_0$-axis,
$(-\sqrt{\vk^2+(m_B-m_\pi)^2}$ to $\sqrt{\vk^2+(m_B-m_\pi)^2}~)$ for Landau cut and 
$(\sqrt{\vk^2+(m_\pi+m_B)^2}$ to $\infty~)$ for unitary cut in positive $k_0$-axis.
These are representing the different kinematic regions  
where the imaginary part of the nucleon self-energy becomes non-zero
because of the different $\delta$ functions in Eq.~(\ref{self_LU}).
The $\Gamma_N$ for all $\pi B$ loops (except the $\pi N$)
are coming from the Landau cut contribution associated with
the third term of Eq.~(\ref{self_LU}), which can be simplified as
\bea
\Gamma_N&=&\frac{1}{16\pi\vk}\int^{\tom^-}_{\tom^+}d\tom
\{n_l(\tom)+n^+_u(\om^N_k+\tom)\}
\nn\\
&&L(l_0=-\tom,{\vec l}=\sqrt{\tom^2-m_\pi^2},k_0=\om^N_k,\vk)~,
\label{gm_int}
\eea
where
$n_l(\tom)=1/\{e^{\beta\tom}-1\}$, $n^+_u(\om^N_k+\tom)=1/\{e^{\beta(\tom+\om^N_k-\mu_N)}+1\}$, 
$\tom^{\pm}=\frac{R^2}{2m_N^2}(-\om^N_k\pm\vk W)$ with 
$W=\sqrt{1-\frac{4m_\pi^2m_N^2}{R^4}}$ and $R^2=m_N^2+m_\pi^2-m_B^2$.

The effective Lagrangian densities for $BN\pi$ interactions are given below~\cite{Leopold}
\bea
{\cal L}&=&\frac{f}{m_\pi}{\ov \psi}_B\gamma^\mu
\left\{
\begin{array}{c}
i\gamma^5 \\
\unit
\end{array}
\right\}
\psi_N\del_\mu\pi + {\rm h.c.}~{\rm for}~J_B^P=\frac{1}{2}^{\pm}~,
\nn\\
{\cal L}&=&\frac{f}{m_\pi}{\ov \psi}^\mu_B
\left\{
\begin{array}{c}
\unit \\
i\gamma^5
\end{array}
\right\}
\psi_N\del_\mu\pi + {\rm h.c.}~{\rm for}~J_B^P=\frac{3}{2}^{\pm}~,
\label{Lag_BNpi}
\eea
where coupling constants $f/m_\pi$ for different baryons have been fixed from
their experimental vacuum widths in $N\pi$ channel.
With the help of the above Lagrangian densities, one can easily
find
\bea
L(k,l)&=&-\left(\frac{f}{m_\pi}\right)^2\ls(\ks-\ls -Pm_B)\ls
~~~~~~{\rm for}~J_B^P=\frac{1}{2}^{\pm}~,
\nn\\
L(k,l)&=&-\left(\frac{f}{m_\pi}\right)^2(\ks-\ls +Pm_B)l_\mu l_\nu
\left\{-g^{\mn}+\frac{1}{3}\gamma^\mu\gamma^\nu
\right.\nn\\
&&\left.+\frac{2}{3m_B^2}(k-l)^\mu(k-l)^\nu
\right.\nn\\
&&\left.+\frac{1}{3m_B}(\gamma^\mu(k-l)^\nu-(k-l)^\mu\gamma^\nu)\right\}
~{\rm for}~J_B^P=\frac{3}{2}^{\pm}~.
\nn\\
\eea
For simplification the coefficients of $\gamma^0$ and $\unit$
are taken as in Ref.~\cite{Ghosh_N} and their addition gives
\bea
L(k,l)&=&-\left(\frac{f}{m_\pi}\right)^2\left\{\left(\frac{R^2}{2}-m_\pi^2
\right)l_0
\right.\nn\\
&&\left.~~~-Pm_\pi^2m_B\right\}
~{\rm for}~~~~~~~~~~J_B^P=\frac{1}{2}^{\pm}~,
\nn\\
L(k,l)&=&-\left(\frac{f}{m_\pi}\right)^2\frac{2}{3m_B^2}
\left\{\left(\frac{R^2}{2}-m_\pi^2\right)^2
\right.\nn\\
&&\left.-m_\pi^2m_B^2\right\}(k_0-l_0+Pm_B)
~~{\rm for}~J_B^P=\frac{3}{2}^{\pm}~.
\nn\\
\eea
The isospin part of the Lagrangian densities are not written
in the Eq.~(\ref{Lag_BNpi}). The isospin structure for 
$J_B^P={\frac{1}{2}}^\pm$ and $J_B^P={\frac{3}{2}}^\pm$
should be ${\ov \psi}{\vec\tau}\cdot{\vec\pi}\psi$ and
${\ov \psi}{\vec T}\cdot{\vec\pi}\psi$ respectively, where
${\vec T}$ is the spin $3/2$ transition operator
and ${\vec \tau}$ is the Pauli operator. This issue is managed by multiplying 
appropriate isospin factors with the expressions of 
corresponding loop diagrams. 
The isospin factor for $\pi N$ or $\pi N^*$ loop is
$I_{N\rightarrow \pi N,N^*}=3$, whereas for the $\pi\Delta$
or $\pi\Delta^*$ loop, $I_{N\rightarrow \pi \Delta,\Delta^*}=2$. 

All baryon resonances have finite vacuum width in $N\pi$
decay channel. The calculations of these decay widths are
very essential in the present work for two reasons.
First is to fix the coupling constants $f/m_\pi$ for different
$BN\pi$ interaction Lagrangian densities and second is 
to include the effect of these baryon widths ($\Gamma_B$)
on the nucleon thermal width $\Gamma_N$. Using the Lagrangian densities, the vacuum
decay width of baryons $B$ for $N\pi$ channel can be
obtained as
\bea
\Gamma_B(m_B)&=&\frac{I_{N^*\rightarrow\pi N}}{2J_B+1}
\left(\frac{f}{m_\pi}\right)^2\frac{|\vp_{cm}|}{2\pi m_B}
[2m_B|\vp_{cm}|^2
\nn\\
&&+m_\pi^2(\om^N_{cm}-Pm_N)]~{\rm for}~J_B^P=\frac{1}{2}^{\pm}~,
\nn\\
\Gamma_B(m_B)&=&\frac{I_{\Delta,\Delta^*\rightarrow\pi N}}{2J_B+1}
\left(\frac{f}{m_\pi}\right)^2\frac{|\vp_{cm}|^3}{3\pi m_B}
\nn\\
&&[\om^N_{cm}+Pm_N]~{\rm for}~~~~~~~J_B^P=\frac{3}{2}^{\pm}~,
\label{Gam_BNpi}
\eea
where $|\vp_{cm}|=\frac{\sqrt{\{m_B^2-(m_N+m_\pi)^2\}\{m_B^2-(m_N-m_\pi)^2\}}}{2m_B}$
and $\om^N_{cm}=\sqrt{|\vp_{cm}|^2+m_N^2}$.
The isospin factors are
$I_{N^*\rightarrow\pi N}=3$ and $I_{\Delta,\Delta^*\rightarrow\pi N}=1$
for the $N\pi$ decay channels of $N^*$ and $\Delta^*$ (or $\Delta$)
respectively.

Now, the $\Gamma_N$ in Eq.~(\ref{gm_int}) can be convoluted 
(see {\it e.g.} Refs.~\cite{S_rho,S_omega}) as
\bea
\Gamma_N(m_B)&=&\frac{1}{N_B}\int^{m_B+2\Gamma_B(m_B)}
_{m_B-2\Gamma_B(m_B)}
dM_BA_B(M_B)\Gamma_N(M_B)~,
\nn\\
N_B&=&\int^{m_B+2\Gamma_B(m_B)}
_{m_B-2\Gamma_B(m_B)}A_B(M_B)~,
\label{gm_BNpi}
\eea
where
\be
A_B(M_B)=\frac{1}{\pi}{\rm Im}\left[\frac{1}{M_B-m_B+i\Gamma_B(M_B)/2}\right]
\ee
is vacuum spectral function of baryons for their vacuum decay width in $N\pi$
channel. Replacing baryon mass $m_B$ by its invariant mass $M_B$ in Eq.~(\ref{Gam_BNpi}),
one can get the off-mass shell expression of $\Gamma_B(M)$. 
The values of coupling constants $f/m_\pi$, which are fixed from
the experimental values of baryon decay width in $N\pi$ channels~\cite{PDG},
are shown in a Table~(I).

\section{Results and discussion}
\begin{figure}
\begin{center}
\includegraphics[scale=0.35]{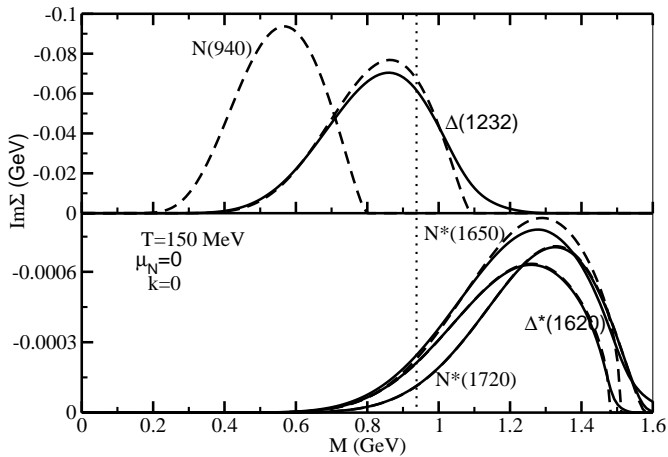}
\caption{Imaginary part of nucleon self-energy for different
$\pi B$ loops are individually shown before (dashed line) and 
after (solid line) folding by corresponding baryon spectral functions.
$B=N(940), \Delta(1232)$ are in upper panel whereas $B=\Delta^*(1620)$,
$N^*(1650), N^*(1720)$ are in lower panel for fixed values
of three momentum of $N$ ($\vk=0$), temperature ($T=0$) and 
baryon chemical potential ($\mu_N=0$).}
\label{ImN_M}
\end{center}
\end{figure}
\begin{figure}
\begin{center}
\includegraphics[scale=0.35]{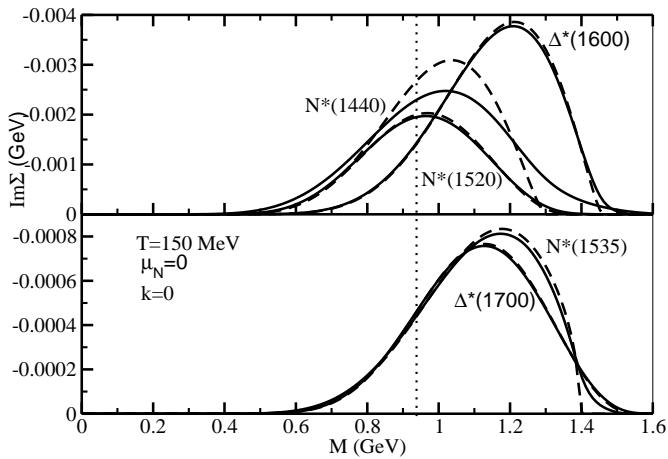}
\caption{Same as Fig.~(\ref{ImN_M}) for the rest of the baryons
$B=N^*(1440)$, $N^*(1520)$, $\Delta^*(1600)$ (upper panel)
and $B=N^*(1535)$, $\Delta^*(1700)$ (lower panel).}
\label{ImN_M2}
\end{center}
\end{figure}
\begin{figure}
\begin{center}
\includegraphics[scale=0.35]{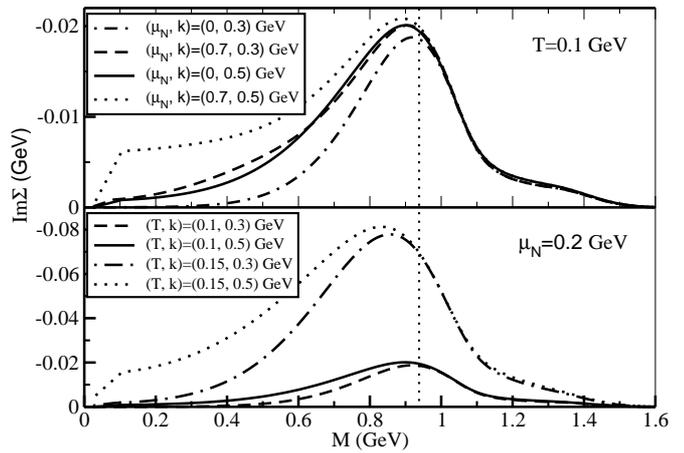}
\caption{Imaginary part of total self-energy for different
sets of nucleon momentum ($\vk$), temperature ($T$) and 
baryon chemical potential ($\mu_N$).}
\label{ImtotN_M2}
\end{center}
\end{figure}
Let us first take a glance at the invariant mass
distribution of imaginary part of nucleon self-energy
for different $\pi B$ loops. Fig.~(\ref{ImN_M})
shows the results for baryons $B=N(940)$, $\Delta(1232)$ (upper panel)
and $B=\Delta^*(1620)$, $N^*(1650)$, $N^*(1720)$ (lower panel), whereas
Fig.~(\ref{ImN_M2}) displays the results for baryons 
$B=N^*(1440)$, $N^*(1520)$, $\Delta^*(1600)$ (upper panel)
and $B=N^*(1535)$, $\Delta^*(1700)$ (lower panel). 
The numerical strengths for $B=N^*(1700)$ and
$N^*(1710)$ are too low to display with the other baryons. 
These results are obtained by replacing $\om^N_k=\sqrt{\vk^2+m_N^2}$ by 
$\om_k=\sqrt{\vk^2+M^2}$ in Eq.~(\ref{gm_int}) (dashed line)
and (\ref{gm_BNpi}) (solid line) for the fixed
values of $\vk=0$, $\mu_N=0$ and $T=0.150$ GeV. 
From the sharp ending of the dashed line, the Landau regions
for different loops are clearly visible. As an example for
$\pi N$ loop the Landau region is $M=0$ to $m_N-m_\pi$, {\it i.e.},
$0$ to $0.8$ GeV. Due to the folding
of the baryon spectral functions, 
these sharp endings are smeared towards higher value of $M$.
Since $\Sigma^R(M)$ also depends on $T$, $\mu_N$ and $\vk$
therefore total contribution of $\Sigma^R(M)$ from all the 
loops has been shown in Fig.~(\ref{ImtotN_M2}) for different
sets of $T$, $\mu_N$ and $\vk$.
\begin{figure}
\begin{center}
\includegraphics[scale=0.35]{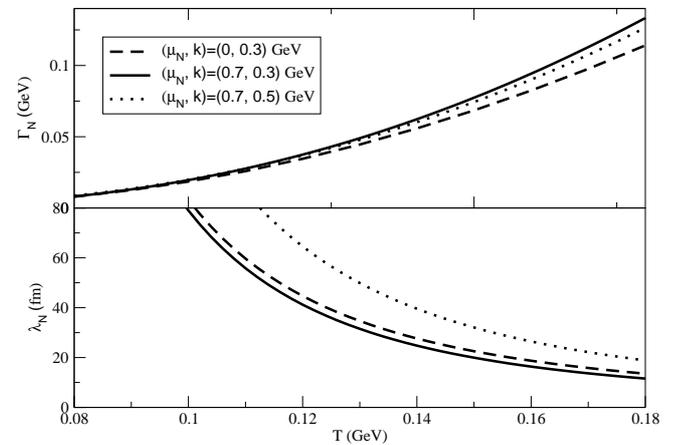}
\caption{The variation of nucleon thermal width $\Gamma_N$
(upper panel) and its corresponding mean free path $\lambda_N$
(lower panel) with $T$ are shown.}
\label{gm_T}
\end{center}
\end{figure}
\begin{figure}
\begin{center}
\includegraphics[scale=0.35]{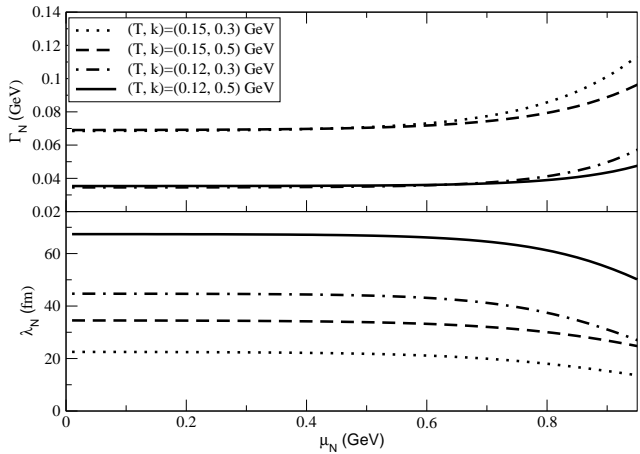}
\caption{The variation of nucleon thermal width $\Gamma_N$
(upper panel) and its corresponding mean free path $\lambda_N$
(lower panel) as a functions of baryon chemical potential 
$\mu_N$ are shown.}
\label{G_mu}
\end{center}
\end{figure}

The nucleon thermal width $\Gamma_N$ is basically 
the contribution of Im$\Sigma^R$ at $M=m_N$, 
which is marked by dotted line. Being an on-shell quantity, $\Gamma_N$ is
associated with the thermodynamical probability of different on-shell 
scattering processes instead of off-shell scattering processes
as described by Weldon for the imaginary part of self-energy in Ref.~\cite{Weldon}. 
Following Weldon's prescription,
forward and inverse scattering of nucleon can be respectively described as follows.
During propagation of $N$, it can disappear by absorbing a thermalized
$\pi$ from the medium to create a thermalized $B$.
Again $N$ can appear by absorbing a thermalized $B$
from the medium as well as by emitting a thermalized $\pi$. The
$n_l(1-n^+_u)$ and $n^+_u(1+n_l)$ are the corresponding statistical
probabilities of the forward and inverse scattering respectively~\cite{Weldon},
because just by adding them, we will get the thermal distribution part 
of Eq.~(\ref{gm_int}), {\it i.e.}, $(n_l+n^+_u)$.

From Eq.~(\ref{gm_int}) or (\ref{gm_BNpi}), we see that 
$\Gamma_N$ depends on temperature $T$, baryon chemical potential $\mu_N$ 
and three momentum $\vk$ of nucleon. 
The upper panels of Fig.~(\ref{gm_T}) and (\ref{G_mu}) are, respectively, displaying
the variation of $\Gamma_N$ with $T$ for different sets of ($\vk,~\mu_N$) 
and of $\Gamma_N$ with $\mu_N$ for different set of ($\vk, T$).
The mean free
path can be defined as $\lambda_N(\vk,T,\mu_N)=\frac{\vk}{\om^N_k \Gamma_N(\vk,T,\mu_N)}$ 
and its corresponding variation with $T$ and $\mu_N$ are respectively shown in the
lower panels of Fig.~(\ref{gm_T}) and (\ref{G_mu}).
The range of $T$ and $\mu_N$, in which $\lambda_N$ is smaller
than the dimension of the medium ($\sim 10-40$ fm, a typical dimension
of strongly interacting matter, produced in the laboratories of HIC),
plays the main role of dissipation via scattering in the medium because
the larger $\lambda$ is associated with the scenario after freeze out of
the medium. From the dashed line of Fig.~(\ref{gm_T}) we see that
$T>0.120$ GeV (but up to $T_c\approx 0.175$ GeV) is that relevant region for baryon 
free nuclear matter ($\mu_N=0$). Whereas for finite baryon
chemical potential (e.g. solid line of Fig.~(\ref{gm_T})
at $\mu_N=0.7$ GeV), this relevant $T$ region will be shifted slightly
toward lower temperature (in addition, $T_c$ is also expected to
decrease with increase of $\mu_N$). Since high momentum ($\vk$)
of constituent particles always helps them to freeze out from the medium,
the relevant $T$ region for nucleon with high $\vk$ is reduced by
shifting towards the high $T$ region. This can be understood by comparing
the solid and dotted lines in the lower panel of Fig.~(\ref{gm_T}).
%
\begin{figure}
\begin{center}
\includegraphics[scale=0.35]{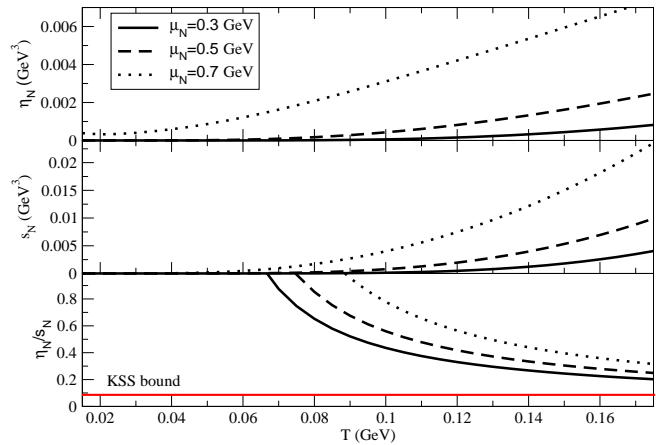}
\caption{(color on-line) The $T$ dependence of $\eta_N$ (upper panel), $s_N$ (middle panel)
and $\eta_N/s_N$ (lower panel) of the nucleonic component.
The straight red line denotes the KSS bound. }
\label{etaN_s_T}
\end{center}
\end{figure}
\begin{figure}
\begin{center}
\includegraphics[scale=0.35]{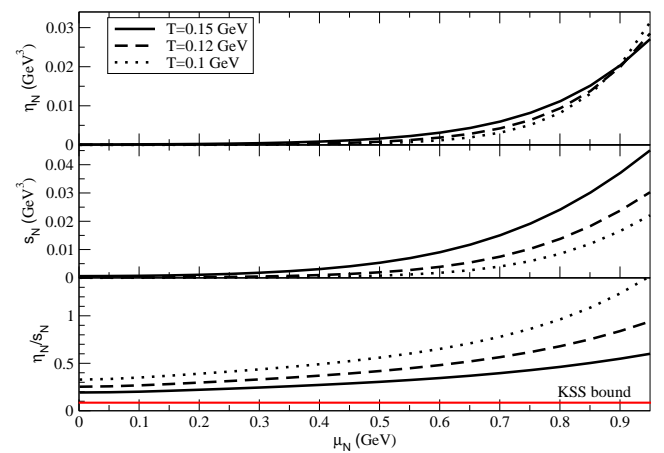}
\caption{(color on line) The variation of $\eta_N$ (upper panel), $s_N$ (middle panel)
and $\eta_N/s_N$ (lower panel) of the nucleonic component 
with $\mu_N$.}
\label{etaN_s_mu}
\end{center}
\end{figure}
\begin{figure}
\begin{center}
\includegraphics[scale=0.35]{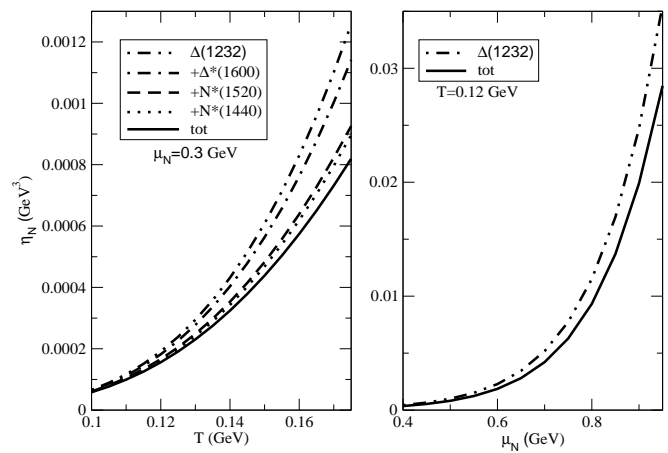}
\caption{The contributions of different $\pi B$ loops in
$\eta_N(T)$ (left panel) and $\eta_N(\mu_N)$ (right panel).}
\label{etaNpiR_T}
\end{center}
\end{figure}

Using the numerical function $\Gamma_N(\vk,T,\mu_N)$ in Eq.~(\ref{eta_last}), 
we get $\eta_N$ as a function of $T$ and $\mu_N$, which are shown in the 
upper panels of Fig.~(\ref{etaN_s_T}) and (\ref{etaN_s_mu}). 
Here we see $\eta_N$ is monotonically increasing with $T$ and $\mu_N$ both.
Using the simple equilibrium expression of entropy density $(s_N)$ for nucleons,
\be
s_N=4\beta\int\frac{d^3\vk }{(2\pi)^3}
\left(\om^N_k+\frac{\vk^2}{3\om^N_k}-\mu_N\right)n^+_k(\om^N_k),
\label{s_N}
\ee
the $\eta_N/s_N$ has been generated as a function
of $T$ and $\mu_N$. From the lower panels of Fig.~(\ref{etaN_s_T}) and (\ref{etaN_s_mu}),
we see that $\eta_N/s_N$ can be reduced by increasing $T$ as well as by decreasing
$\mu_N$. 

In the left and right panels of Fig.~(\ref{etaNpiR_T}), the contributions 
of different loops (dominating loops only) are individually shown
in $\eta_N$ vs $T$ and $\eta_N$ vs $\mu_N$ graphs respectively. The $\pi\Delta$
loop plays a leading role
to generate the typical values ($0.0001-0.01$ GeV$^3$) of $\eta_N$ for
strongly interacting matter because the major part of
the nucleon thermal width is coming from this loop only.
\begin{figure}
\begin{center}
\includegraphics[scale=0.35]{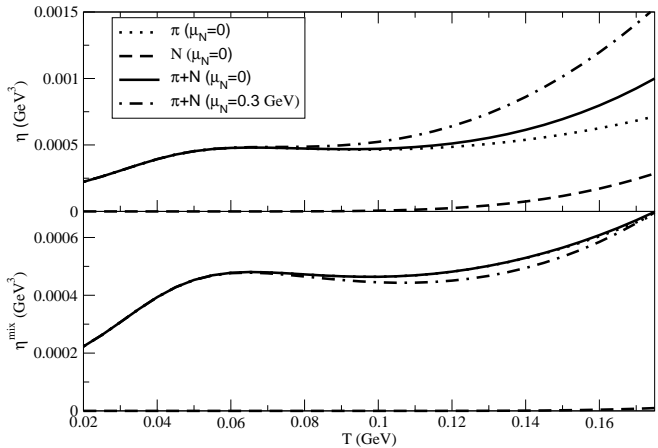}
\caption{The $T$ dependence of shear viscosity for pionic (dotted line),
nucleonic (dashed line) components and their total at $\mu_N=0$ 
(solid line) and $\mu_N=0.3$ GeV (dash-dotted line). The upper
and lower panel contain the results without and with mixing effect,
obtained from Eq.(\ref{eta_tot}) and (\ref{etamix_tot}) respectively.}
\label{etamix_T}
\end{center}
\end{figure}
\begin{figure}
\begin{center}
\includegraphics[scale=0.35]{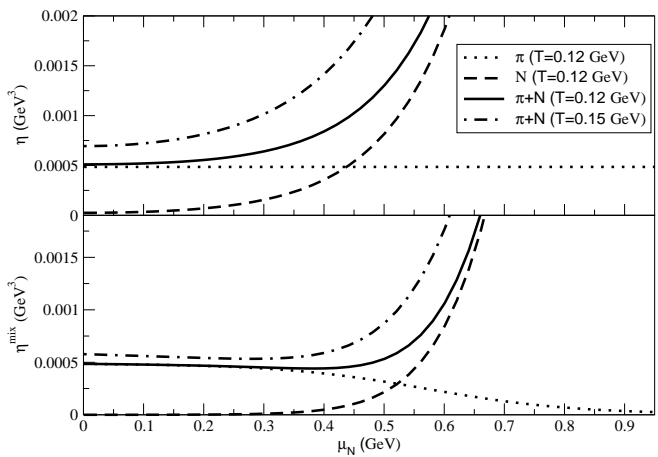}
\caption{Corresponding results of Fig.~(\ref{etamix_T})
against the $\mu_N$ axis with two different temperatures.}
\label{etamix_mu}
\end{center}
\end{figure}

Up to now, we have calculated the contribution of shear viscosity from
nucleon thermal width, although a major contribution comes
from the thermal width of pion. Hence, one should add the pionic contribution
with nucleon contribution for getting total shear viscosity of
nuclear matter at finite temperature and density.
In our recent work~\cite{GKS}, the shear viscosity, coming from pionic thermal width
has already been addressed. The one-loop Kubo expression of shear
viscosity and ideal expression of entropy density for pionic components
are respectively given below,
\be
\eta_\pi = \frac{\beta}{10\pi^2}\int\frac{d^3k\,\vk^6}{\Gamma_{\pi}{\om^{\pi}_k}^{2} } 
\, n_k(\om^\pi_k)\left[1+n_k(\om^\pi_k)\right]~,
\label{eta1_final}
\ee
and
\be
s_\pi = 3\beta\int \frac{d^3\vk}{(2\pi)^3} \left(\om^\pi_k+\frac{\vk^2}{3\om^\pi_k}\right)
n_k(\om^\pi_k)~,
\label{s_pi}
\ee
where $n_k(\om^\pi_k) =1/\{e^{\beta\om^\pi_k}-1\}$
is the Bose-Einstein distribution function of pion with 
$\om^\pi_k = (\vk^2 +m_\pi^2)^{1/2}$, and $\Gamma_\pi$ is the 
thermal width of $\pi$ mesons in the medium due to $\pi\sigma$
and $\pi\rho$ fluctuations.

Now, adding that pion contribution with the 
nucleon, one can simply get the total shear viscosity of nuclear matter as
\be
\eta_{\rm tot}=\eta_\pi +\eta_N~,
\label{eta_tot}
\ee
where $\eta_\pi$ and $\eta_N$ do not face any mixing effect of 
pion density, $\rho_\pi=3\int\frac{d^3k}{(2\pi)^3}n_k(\om_k^\pi)$ and
nucleon density, $\rho_N=4\int\frac{d^3k}{(2\pi)^3}n^+_k(\om_k^N)$.
However, viscosity of single component gas should be different from
the viscosity of that component in a mixed gas~\cite{Itakura,mix}. To
incorporate this mixing effect for rough estimation, we follow
the approximated relation~\cite{Itakura,mix}
\be
\eta^{\rm mix}_{\rm tot}=\eta^{\rm mix}_\pi +\eta^{\rm mix}_N~,
\label{etamix_tot}
\ee
where 
\be
\eta^{\rm mix}_\pi=\frac{\eta_\pi}{1+\left(\frac{\rho_N}{\rho_\pi}\right)
\left(\frac{\sigma_{\pi N}}{\sigma_{\pi\pi}}\right)\sqrt{\frac{1+m_\pi/m_N}{2}}}
\label{etamix_pi}
\ee
and
\be
\eta^{\rm mix}_N=\frac{\eta_N}{1+\left(\frac{\rho_\pi}{\rho_N}\right)
\left(\frac{\sigma_{\pi N}}{\sigma_{NN}}\right)\sqrt{\frac{1+m_N/m_\pi}{2}}}~.
\label{etamix_N}
\ee
For simplicity, the cross sections of all kinds of scattering are taken as constant
with same order of magnitude ({\it i.e.} 
$\sigma_{\pi\pi}\approx \sigma_{\pi N} \approx \sigma_{NN}$).
In the upper panels of Fig.~(\ref{etamix_T}) and (\ref{etamix_mu}), the $T$ and
$\mu_N$ dependence of $\eta_\pi$ (dotted line),
$\eta_N$ (dashed line) and their total $\eta_{\rm tot}$ (solid line and 
dash-dotted line for two different values of $\mu_N$ and $T$) are separately shown.
Whereas lower panel of the figures show their corresponding mixing 
effect following from Eq.~(\ref{etamix_pi}), (\ref{etamix_N}) and (\ref{etamix_tot}).
From the Fig.~(\ref{etamix_mu}), one should notice that the independent
nature of $\eta_\pi(\mu_N)$ has been changed to a decreasing function 
due to mixing effect. 
Similar qualitative trend has been seen in Ref.~\cite{Itakura}.
\begin{figure}
\begin{center}
\includegraphics[scale=0.35]{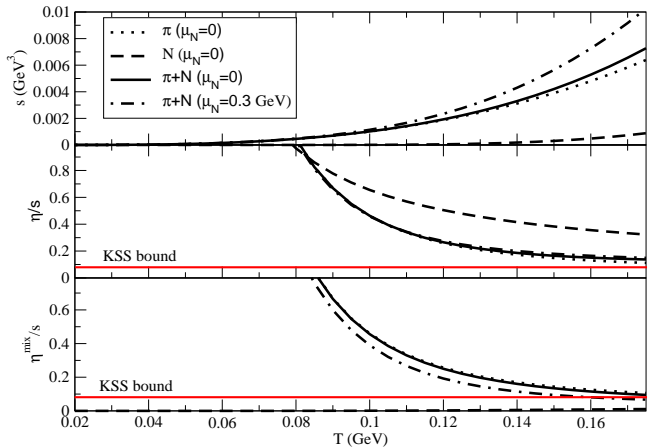}
\caption{(color on-line) The entropy density (upper panel), viscosity
to entropy density ratio without (middle panel) and with
(lower panel) mixing effect as functions
of $T$.}
\label{etamix_s_T}
\end{center}
\end{figure}
\begin{figure}
\begin{center}
\includegraphics[scale=0.35]{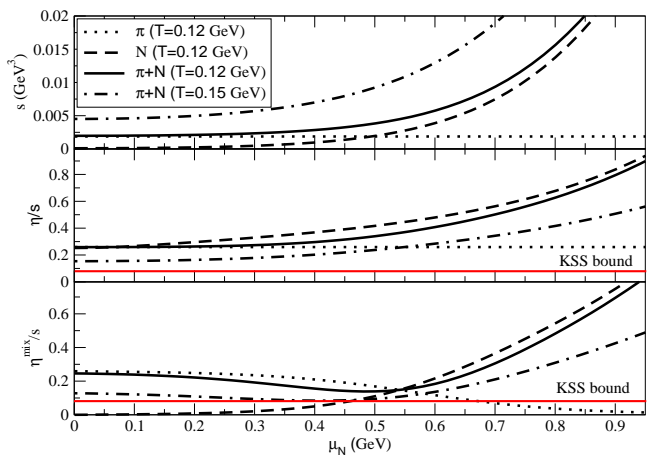}
\caption{(color on-line) The corresponding results of Fig.~(\ref{etamix_s_T}) are
shown with respect to $\mu_N$.}
\label{etamix_s_mu}
\end{center}
\end{figure}

The entropy density of nucleon component from Eq.~(\ref{s_N}), pion
component from Eq.~(\ref{s_pi}) and their total $s_{\rm tot}=s_N+s_\pi$
are individually shown in the upper panels of Fig.~(\ref{etamix_s_T})
and (\ref{etamix_s_mu}) as functions of $T$ and $\mu_N$ respectively.
The corresponding $\eta/s$ without (middle panel) and with (lower panel) 
mixing effect are shown in Fig.~(\ref{etamix_s_T})
and (\ref{etamix_s_mu}) as a function of $T$ and $\mu_N$ respectively.
The decreasing nature of total $\eta/s(T)$ qualitatively remains the same after
incorporating the mixing effect whereas an increasing function of the total
$\eta/s(\mu_N)$ transforms to a decreasing function due to this mixing effect.
Comparing our results with the results of Itakura et al.~\cite{Itakura}, 
where $\eta/s(\mu_N)$ also reduces with $\mu_N$, the mixing effect appears to be very
important. However, the total $\eta/s(\mu_N)$ in mixing scenario becomes an 
increasing function beyond $\mu_N\approx0.5$ GeV because the increasing
rate of $\eta^{\rm mix}_N(\mu_N)$ dominates over the decreasing rate
of $\eta^{\rm mix}_\pi(\mu_N)$ in that region. Using the effective hadronic
Lagrangian, the conclusion of our results should be concentrated within 
regions of $0.100$ GeV $<T<0.160$ GeV and $0<\mu_N<0.500$ GeV.

\section{Summary and Conclusion}
Owing to the Kubo relation, the shear viscosity can be 
expressed in terms of two point function of the viscous
stress tensors at finite temperature. 
By using the real-time thermal field theoretical
method, this two point function has been represented as $NN$ loop diagram
when the nucleons are considered as constituent particles of the medium.
A finite nucleon thermal width $\Gamma_N$ has been traditionally
included in the nucleon propagators of the $NN$ loop for getting a 
non-divergent shear viscosity $\eta_N$. This nucleon thermal width is obtained
from the one-loop self-energy of nucleon at finite temperature
and density. Different possible pion baryon loops are accounted
to calculate the total $\Gamma_N$, which depends on the
three momentum of nucleons ($\vk$) as well as the medium
parameters $T$ and $\mu_N$. Using the numerical function
$\Gamma_N(\vk,T,\mu_N)$, $\eta_N$ and $\eta_N/s_N$
are numerically generated as functions of $T$ and $\mu_N$.
Adding the pionic contribution taken from Ref.~\cite{GKS} with the numerical
values of the nucleonic component, we have obtained the total shear
viscosity, where a gross mixing effect of two component system
has been implemented. Along the temperature axis, the shear viscosity
of both pion and nucleon components appear as increasing
function, whereas along the $\mu_N$ axis shear viscosity of pion
component changes from its constant behavior to a decreasing
function due to presence of mixing effect.
The total shear viscosity to entropy density ratio 
($\eta^{\rm mix}_{\rm tot}/s_{\rm tot}$)
for the pion-nucleon mixed gas
reduces with increasing $T$ as well as $\mu_N$
and quantitatively becomes very close to the KSS
bound.
This behavior indicates that $\eta^{\rm mix}_{\rm tot}/s_{\rm tot}$ tends
to reach its minimum value near the transition temperature
at vanishing as well as finite value of $\mu_N$.
According to these results, the finite
baryon chemical potential helps the nuclear matter to come closer
to its (nearly) perfect fluid nature.

{\bf Acknowledgment :} The work is financially supported
by Fundacao de Amparo a Pesquisa do Estado de Sao Paulo, 
FAPESP (Brazilian agencies) under Contract No. 2012/16766-0.
I am very grateful to Prof. Gastao Krein for his academic
and non-academic support during my postdoctoral period in Brazil.
I would also like to thank Abhishek Mishra, Sandeep Gautam and 
Supriya Mondal for their useful help
while writings this article.

\end{document}